\documentclass{ctr}

\usepackage{ctrfont}
\usepackage{natbib}

\usepackage{url}
\usepackage{amsmath}
\usepackage{amssymb}

\usepackage{graphicx}

\usepackage{psfrag}
\usepackage{epsfig}

\newcommand{\p}{\partial}

\title{Estimated aleatoric uncertainty from initial and inlet conditions for variable density mixing}
\shorttitle{Variable density flow IC sensitivity}
\author{J.~F. Heyse, Z. Huang \and G. Iaccarino\footnote{CTR, Stanford University}}
\shortauthor{Heyse et.~al.}

\begin{document}


\maketitle

\section{Motivation and objectives} 

The physics of fluid mechanics problems are governed by partial differential equations describing the conservation of mass (then called continuity equation), momentum (Navier-Stokes equations), and energy, as well as algebraic equations such as equations of state, which relate state variables. 
Fluid mechanics problems of engineering interest are often too complex to computationally solve the corresponding systems of coupled equations resolving all spatial and temporal flow scales as is done in Direct Numerical Simulations (DNS), or to solve filtered equations that partly resolve and partly model the energy-containing scales as is done in Large Eddy Simulations (LES).
Instead, many industrial problems are solved using Reynolds-averaged Navier-Stokes (RANS) equations, which are computationally much cheaper than LES or DNS. Through the ensemble-averaging there appear correlations in the RANS equations between fluctuating velocity components, known as Reynolds stresses, or between fluctuating scalar variables and velocity components.
No explicit solutions are available for these correlations, which therefore need to be modeled by so-called turbulence models.

Given the inherent manifold limitations of numerical simulations, and specifically RANS simulations, at representing engineering problems of interest, it is important to find ways to quantify the uncertainty in the predictions. Without knowledge about the uncertainties, there cannot be trust in model predictions. 
Accuracy and robustness, which in this context means a low sensitivity to variations in the input, are important to the assessment of the prediction quality.
Identifying and quantifying sources of uncertainty can be used to develop strategies to reduce that uncertainty \citep{gurkan-2011} or to direct experimental work towards identified key measurement parameters \citep{andrews-2013}.

Uncertainty is commonly classified into aleatoric and epistemic sources. 
Aleatoric uncertainty is attributed to stochastic behavior that cannot be known, but will lead to inherent variability when repeating the same experiment multiple times.
\cite{granado-ortiz-2018} studied such parametric uncertainty in the computation of an underexpanded high-speed jet, treating the stagnation pressure $p_{s}$ and turbulent-to-laminar viscosity ratio $R_{t}$ as stochastic variables. First, they compared generalized polynomial chaos with Kriging surrogates as non-intrusive methods of uncertainty quantification. Second, they analyzed the sensitivity of their RANS predictions to the stochastic variables locally at different regions in the domain.
Epistemic uncertainty, on the other hand, is attributed to missing knowledge about something that one could in principle know.
In the case of RANS simulations, such structural uncertainty is associated with the turbulence model.
All these models are limited because they are not always able to represent all of the physics relevant to the problem, and some models may not adequately capture initial or boundary conditions.
\cite{emory-2013} introduced perturbations of the eigenvalues of the Reynolds stress predictions from turbulence models towards three different limiting states in a barycentric map. The range of the solutions is used as an estimate of the model-form uncertainty.
\cite{iaccarino-2017} added perturbations of the eigenvectors to this framework, which in this extended form is referred to as eigenspace perturbations.

The objective of this report is to study the aleatoric uncertainty of two different variable density flows to its inlet and initial conditions, respectively.
Flows with variable density instabilities are commonly characterized by the Atwood number $A=(\rho_{1}-\rho_{2})/(\rho_{1}+\rho_{2})$, which can be interpreted as a non-dimensional density ratio, where $\rho_{1}$ and $\rho_{2}$ are the densities of the heavier and lighter fluid, respectively.
These flows occur in a variety of different systems with a wide range of scales, from astrophysics to atmospheric flows to inertial confinement fusion or reacting flows. 
The two cases studied in this report are the turbulent mixing of a jet in a co-flow at small $A$, and the Rayleigh-Taylor (RT) mixing in a tilted rocket rig at medium $A$. 
The jet in a co-flow involves the spatial evolution of a flow with significant anisotropies, and the tilted rocket rig case incorporates significant 2d effects on the Rayleigh-Taylor mixing region, which makes both of them useful in the context of assessing turbulence models \citep{charonko-2017,denissen-2014}.
For a meaningful assessment it is important to know the uncertainty of the reference data associated with initial and boundary conditions. \cite{boersma-1998} used DNS to study the influence of the shape of the inlet velocity profile on far-field velocity and shear stress profiles. \cite{andrews-2013} computed normalized sensitivity indices for 1d Rayleigh Taylor mixing, initializing the RANS simulations using a mixing layer between the two fluids.
In this report, estimates of the uncertainty due to the inlet and initial conditions, respectively, are made for the two mentioned cases and compared to reference data.





\section{Methodology}

To estimate that initial and inlet uncertainty, parameters of those conditions are varied according to estimated uncertainty levels from reference experiments. 
Estimated distributions of measurement uncertainties from an experimental report are used to do Monte-Carlo sampling for some inlet parameters of the first case. 
For the second case, a perturbation to the initial interface location is varied over an expected range of parameter values. 
Estimates of the uncertainty of the results are made for both cases and compared against reference experiments.

\subsection{Jet in co-flow case}

The first case considered is the turbulent jet in a co-flow. \cite{charonko-2017} experimentally studied the mixing of a turbulent jet in a co-flow of lighter density making PIV and PLIF measurements. A vertical turbulent mixing tunnel of square cross section was used, with the jet moving downward, entering through a supply tube placed at the center of the cross section. 
The measurements were taken at two different Atwood numbers, $A=0.09$ and $A=0.62$.
The velocity uncertainties were estimated to be on average up to $1.5\%$ of full scale, and the density uncertainty to be up to $4\%$. These uncertainties correspond to the bounds of the $95\%$ interval for assumed normal distributions \citep{charonko-2013}.

Based on the $A=0.09$ experiments, simulations are done to study the sensitivity of the mixing to the inlet conditions of the turbulent jet. 
The side walls are not expected to have an important effect on the flow in the region of interest around the jet inlet, and so the axi-asymmetry is expected to be insignificant.
The setup in the simulations is therefore approximated as axi-symmetric, with an outer-domain diameter corresponding to the side length of the square cross section from the experiments. 
The 2d mesh has a resolution of $150\times44$ cells in the vertical and radial directions. 

Baseline values for inlet velocities and densities as well as the inlet turbulent kinetic energy are adapted from the experiment parameters and measurements, and the inlet turbulent length scale is adapted from \cite{brown-2015}.
The simulations are initialized with undisturbed co-flow everywhere, and then they are run for $2.8$ run-through times of the co-flow, by which time they have converged to steady state.

\begin{figure}
\centering
\begin{center}
\includegraphics[width=0.6\textwidth]{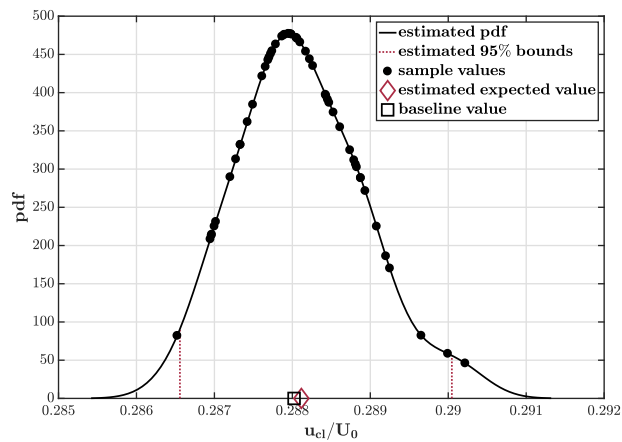}
\caption{Illustration of estimated pdf of normalized centerline velocity at $x/d_{0}=9.39$ for $1.5\%$ $k_{0}$ perturbation, including sample data points, estimated $95\%$ bounds, estimated expected value, and baseline value.}
\label{fig:pdf-estimate}
\end{center}
\end{figure}

The quantitative estimates of the uncertainty of velocity and density measurements in the reference experiments allow us to sample from those assumed distributions to obtain distributions on the model predictions.
The sensitivity to the inlet conditions is therefore studied by sampling values of inlet velocity $U_{0}$, inlet turbulent kinetic energy $k_{0}$, and inlet density $\rho_{0}$. For $k_{0}$, the corresponding turbulent velocity 
\begin{equation}
    u_{k_{0}} = \sqrt{2k_{0}}
\end{equation}
is sampled. The estimated variances for velocity $\sigma_{U_{0}}^{2}$ and density $\sigma_{\rho_{0}}^{2}$ are used, as well as the baseline values for the expected values. The inlet parameters are thus sampled as
\begin{equation}
\begin{split}
U_{0} &\sim \mathcal{N}(\mu_{U_{0}},\,\sigma_{U_{0}}^{2})\;, \\
u_{k_{0}} &\sim \mathcal{N}(\sqrt{2\mu_{k_{0}}},\,\sigma_{U_{0}}^{2})\;, \\
\rho_{0} &\sim \mathcal{N}(\mu_{\rho_{0}},\,\sigma_{\rho_{0}}^{2})\;.
\end{split}
\end{equation}
The three inlet parameters are sampled individually to determine the sensitivity of the steady-state normalized centerline velocity and normalized centerline turbulent kinetic energy profiles to the respective parameters. 
The velocity is normalized by the inlet velocity, while the turbulent kinetic energy is normalized as
\begin{equation}
    \tilde{k}^{*} = \left( \frac{2/3 \tilde{k}}{\tilde{u}_{cl}^{2}-\bar{u}_{\infty}^{2}} \right)^{1/2} \;.
\end{equation}
Fifty-two Monte-Carlo samples are made for each parameter. The probability density function (pdf) of the predictions is then estimated using the Matlab function \texttt{ksdensity}, which bases the estimates on a normal kernel smoothing function.
Figure \ref{fig:pdf-estimate} shows the estimated pdf for the normalized centerline velocity at $x/d_{0}=9.39$ with $1.5\%$ $k_{0}$ perturbation.
This analysis is repeated with an increased inlet parameter uncertainty of $25\%$.

\subsection{Tilted rocket rig case}

\begin{figure}
\centering
\begin{center}
\includegraphics[width=\textwidth]{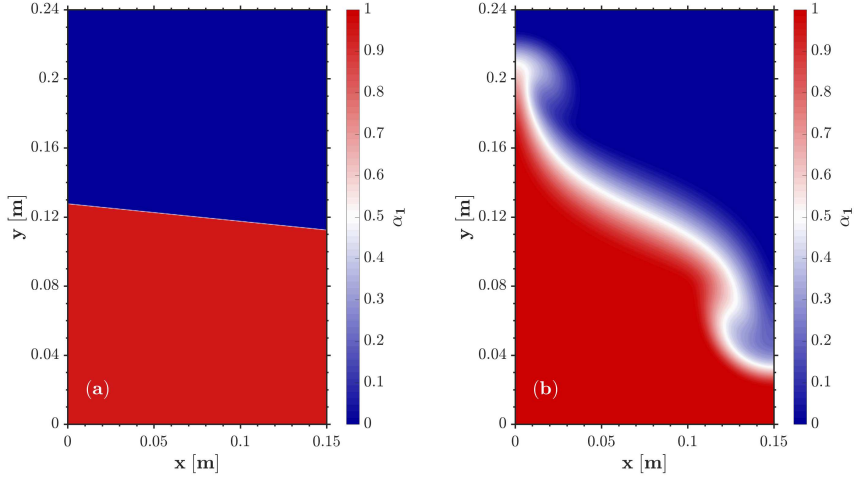}
\caption{Tilted rocket rig volume fraction of the heavy fluid $\alpha_{1}$ without initial perturbation. (a) at $\tau=0.033$, (b) at $\tau=1.827$.}
\label{fig:sample-rocket-rig}
\end{center}
\end{figure}

The second case considered is the tilted rocket rig, originally designed at the Atomic Weapons Establishment to study the Rayleigh Taylor instability \citep{youngs-1989}. A capsule containing two fluids is accelerated downwards by small rocket engines. The heavier of the two fluids is initially at the bottom and is pushed by the acceleration into the lighter fluid. The initial interface is tilted at some angle $\theta$. Bubble height $H_{b}$ and spike height $H_{s}$ are two of the measures used to describe the growth of the Rayleigh-Taylor instability. 
They are defined as the most extreme vertical location of the light fluid to penetrate into the heavy fluid and the heavy fluid into the light fluid, respectively.
Figure \ref{fig:sample-rocket-rig} shows the volume fraction of the heavy fluid $\alpha_{1}$ at some very early time and at some later time from a numerical simulation without initial interface perturbation. The plot at the later time shows the penetration of the spike on the left and of the bubble on the right.

The configuration studied in this report is case $110$ from \cite{smeeton-1987}. The tank has height $H=240$ mm and width $W=150$ mm, and the initial interface angle is $\theta=5.77^{\circ}$. 
The acceleration is $35$ times standard gravity, and the Atwood number is $A=0.482$. 
For these experiments, no detailed initial conditions of the interface and therefore no uncertainties associated with the initial interface were described. However, long-wavelength initial perturbations in the experiments were reported to affect the growth rates of bubble and spike penetration \citep{denissen-2014, youngs-2009}, and \cite{kucherenko-1996} and \cite{dimonte-2000} both reported random initial perturbation amplitudes of $0.4$ mm for similar RT experiments.

Thus, long-wavelength initial interface perturbations are used for the numerical studies in this report. 
Since there is no information available on the distribution of the initial interface conditions, no distribution of the model predictions can be obtained from Monte-Carlo sampling. 
Instead, the wavelength is varied from $W/15$ to $W/1$ in integer steps of the denominator, leading to $15$ different wavelengths.
The range of bubble and spike height values over the 15 different wavelengths is used as a measure of the uncertainty due to the initial condition.
This uncertainty estimate is done for three different perturbation amplitudes. The nominal perturbation amplitude is $0.4$ mm, but simulations with amplitudes $0.2$ mm and $0.6$ mm are run as well.

The simulations are 2d and have a uniform grid of $640\times400$ cells in the vertical and horizontal directions.

\subsection{RANS simulations}

All simulations were carried out using OpenFOAM v1812 with a solver for two compressible fluids using volume-fraction-based interface capturing. 
Spatial discretization is second order, except for divergence computations which are first order upwind. 
The implicit Euler method is used for time advancement.

The Besnard-Harlow-Rauenzahn model is a single-fluid turbulence model, developed specifically for variable density flows \citep{besnard-1992}. In the original variant of this model, BHR-1, transport equations for the turbulent kinetic energy $k$, a turbulent length scale $S$, and a turbulent mass flux velocity $a_{i}=-\overline{u_{i}^{\prime\prime}}=\tilde{u}_{i}-\bar{u}_{i}$ are solved. These transport equations are
\begin{equation}
\begin{split}
    \frac{\p \bar{\rho} k}{\p t} + \frac{\p \bar{\rho} k \tilde{u}_{j}}{\p x_{j}}
    =& a_{j} \frac{\p \bar{p}}{\p x_{j}} - R_{ij} \frac{\p \tilde{u}_{i}}{\p x_{j}}
    + \frac{\p}{\p x_{j}} \left[ \bar{\rho} \nu_{T} \frac{\p k}{\p x_{j}} \right]
    - \bar{\rho} \frac{k^{3/2}}{S}\;, \\
    \frac{\p \bar{\rho} S}{\p t} + \frac{\p \bar{\rho} S \tilde{u}_{j}}{\p x_{j}}
    =& \frac{S}{k} \left[ \left( \frac{3}{2}-C_{4} \right) a_{j} \frac{\p \bar{p}}{\p x_{j}} - \left( \frac{3}{2}-C_{1} \right) R_{ij} \frac{\p \tilde{u}_{i}}{\p x_{j}} \right] \\
    & - C_{3} \bar{\rho} S \frac{\p \tilde{u}_{j}}{\p x_{j}} 
    + \frac{\p}{\p x_{j}} \left( \frac{\bar{\rho} \nu_{T}}{\sigma_{\epsilon}} \frac{\p S}{\p x_{j}} \right)
    - \left( \frac{3}{2} - C_{2} \right) \bar{\rho} k^{1/2}\;, \\
    \frac{\p \bar{\rho} a_{i}}{\p t} + \frac{\p \bar{\rho} a_{i} \tilde{u}_{j}}{\p x_{j}}
    =& C_{a2} b \frac{\p \bar{p}}{\p x_{i}} - \frac{R_{ij}}{\bar{\rho}} \frac{\p \bar{\rho}}{\p x_{j}}
    -\frac{C_{a1} \bar{\rho} a_{i} k^{1/2}}{S} - \bar{\rho} a_{j} \frac{\p \tilde{u}_{i}}{\p x_{j}}\;.
\end{split}
\label{eq:bhr-transport}
\end{equation}
$R_{ij}$ is the Reynolds stress, $\nu_{T}$ is the turbulent viscosity, and $C_{i}$ are model coefficients.
The density self-correlation 
\begin{equation}
    b = -\overline{\rho^{\prime} \left(\frac{1}{\rho}\right)^{\prime}}
\end{equation}
is approximated by the algebraic equation
\begin{equation}
    b = C_{b} \alpha_{0} \alpha_{1}\frac{(\rho_{1}-\rho_{0})^{2}}{\rho_{0}\rho_{1}}\;,
\end{equation}
where $\alpha_{i}$ is the volume fraction and the subscripts $0$ and $1$ denote the light and heavy fluid.

\section{Results}

\subsection{Jet in co-flow case}

\begin{figure}
\centering
\begin{center}
\includegraphics[width=\textwidth]{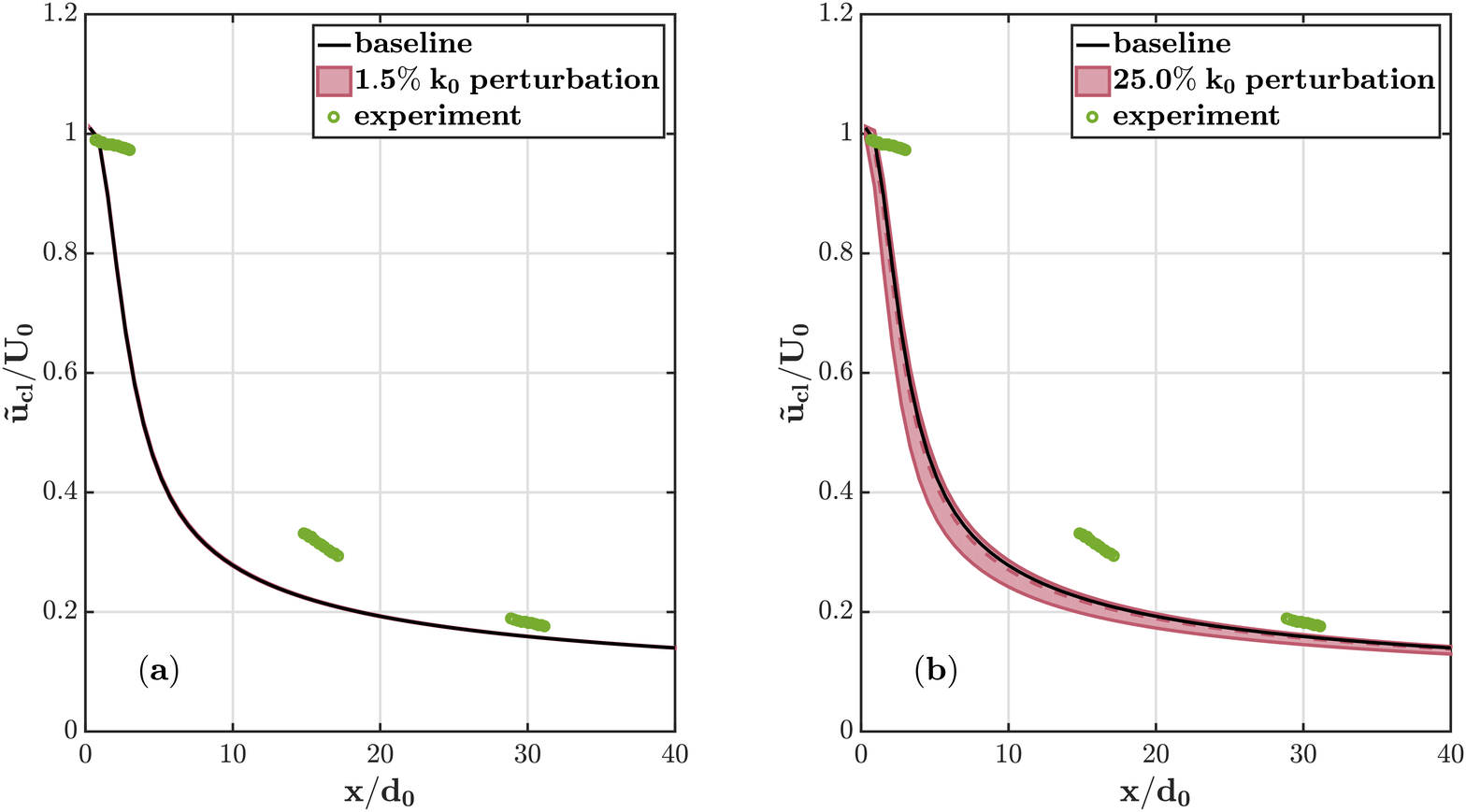}
\caption{Normalized centerline velocity profile with baseline, estimated uncertainty to $k$ inlet condition, and experimental data. (a) with $1.5\%$ perturbation, (b) with $25.0\%$ perturbation.}
\label{fig:jet-cl-k}
\end{center}
\end{figure}

\begin{figure}
\centering
\begin{center}
\includegraphics[width=\textwidth]{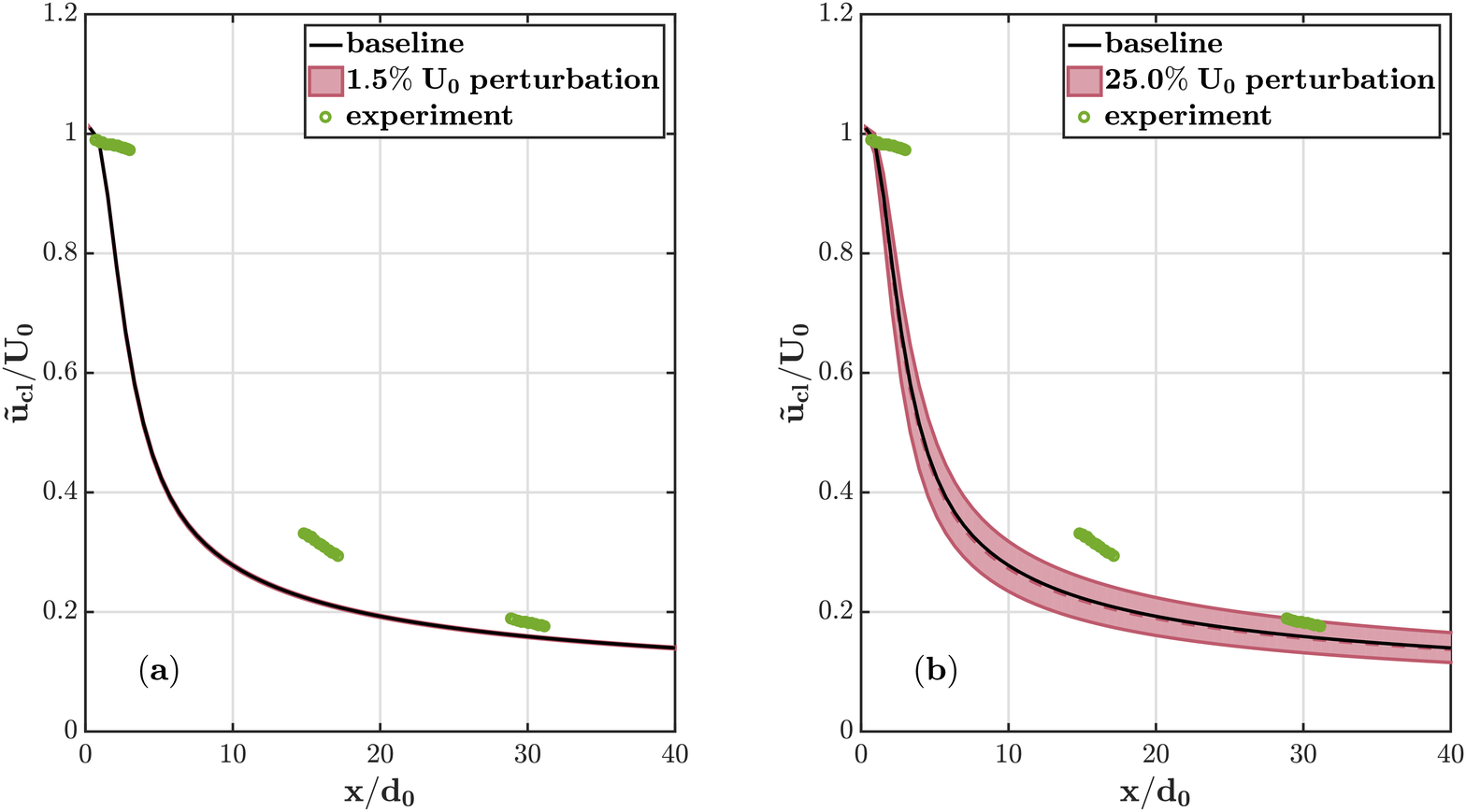}
\caption{Normalized centerline velocity profile with baseline, estimated uncertainty to $U$ inlet condition, and experimental data. (a) with $1.5\%$ perturbation, (b) with $25.0\%$ perturbation.}
\label{fig:jet-cl-U}
\end{center}
\end{figure}

\begin{figure}
\centering
\begin{center}
\includegraphics[width=\textwidth]{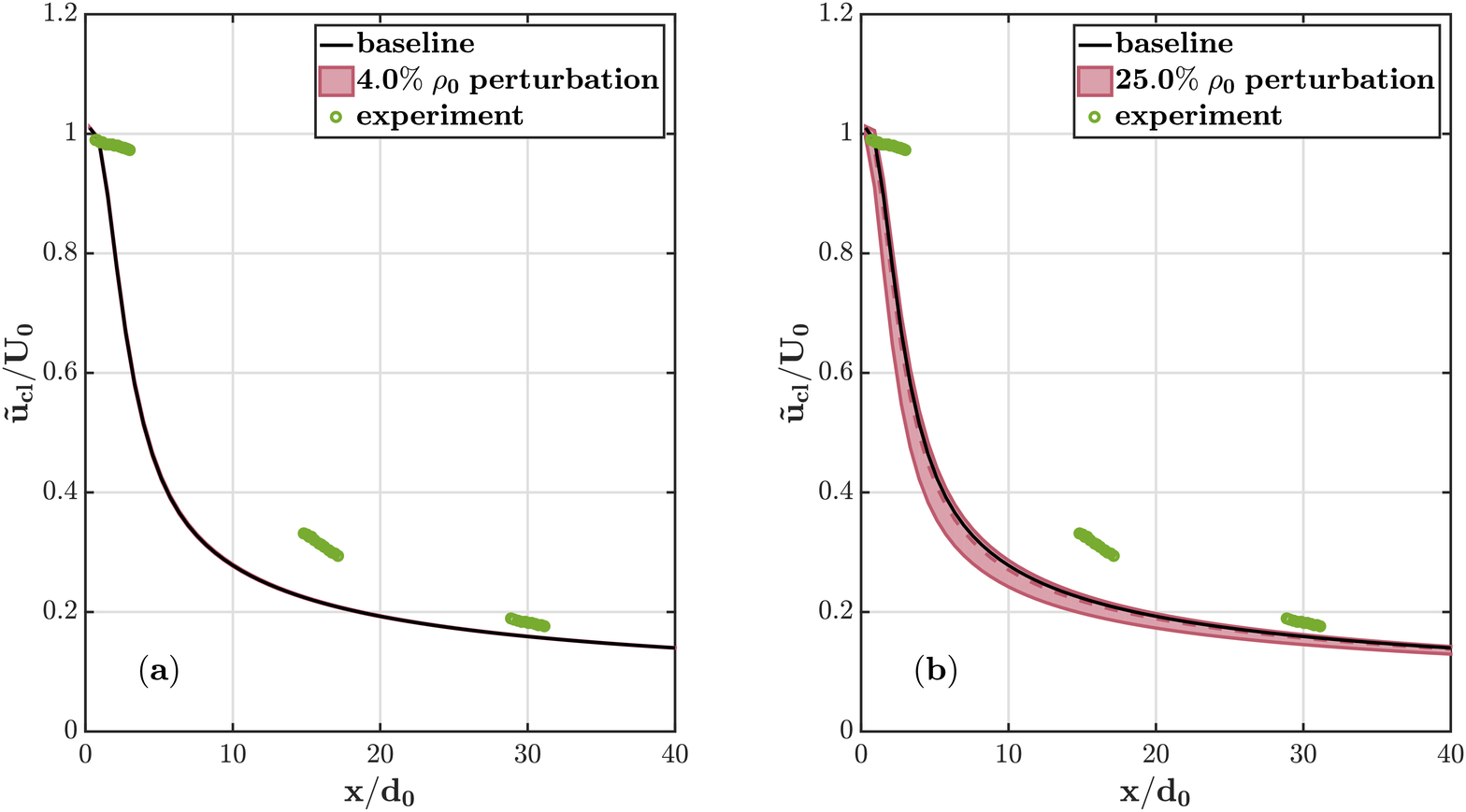}
\caption{Normalized centerline velocity profile with baseline, estimated uncertainty to $\rho$ inlet condition, and experimental data. (a) with $4.0\%$ perturbation, (b) with $25.0\%$ perturbation.}
\label{fig:jet-cl-r}
\end{center}
\end{figure}

\begin{figure}
\centering
\begin{center}
\includegraphics[width=\textwidth]{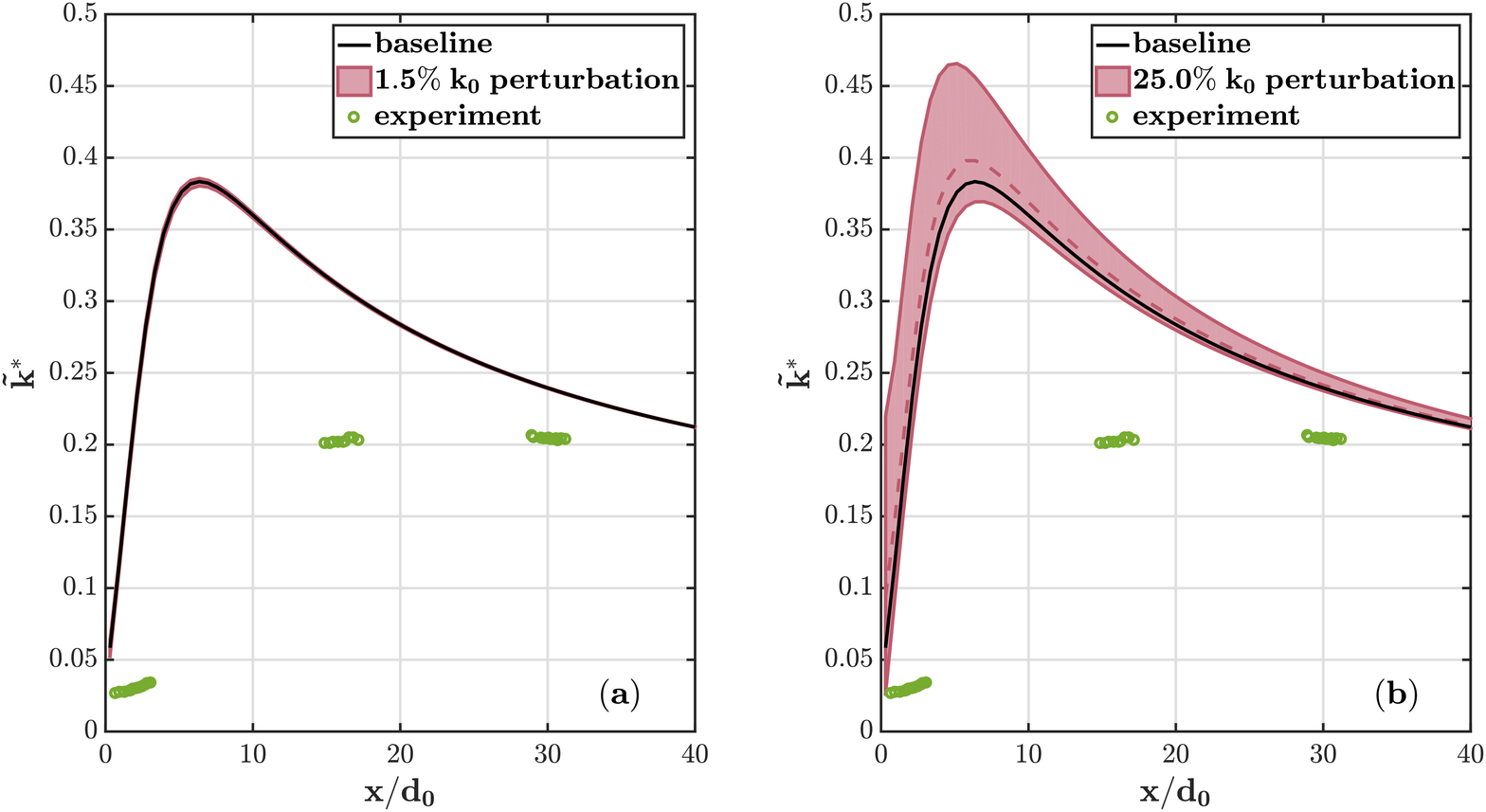}
\caption{Normalized centerline turbulent kinetic energy profile with baseline, estimated uncertainty to $k$ inlet condition, and experimental data. (a) with $1.5\%$ perturbation, (b) with $25.0\%$ perturbation.}
\label{fig:jet-cl-k-k}
\end{center}
\end{figure}

\begin{figure}
\centering
\begin{center}
\includegraphics[width=\textwidth]{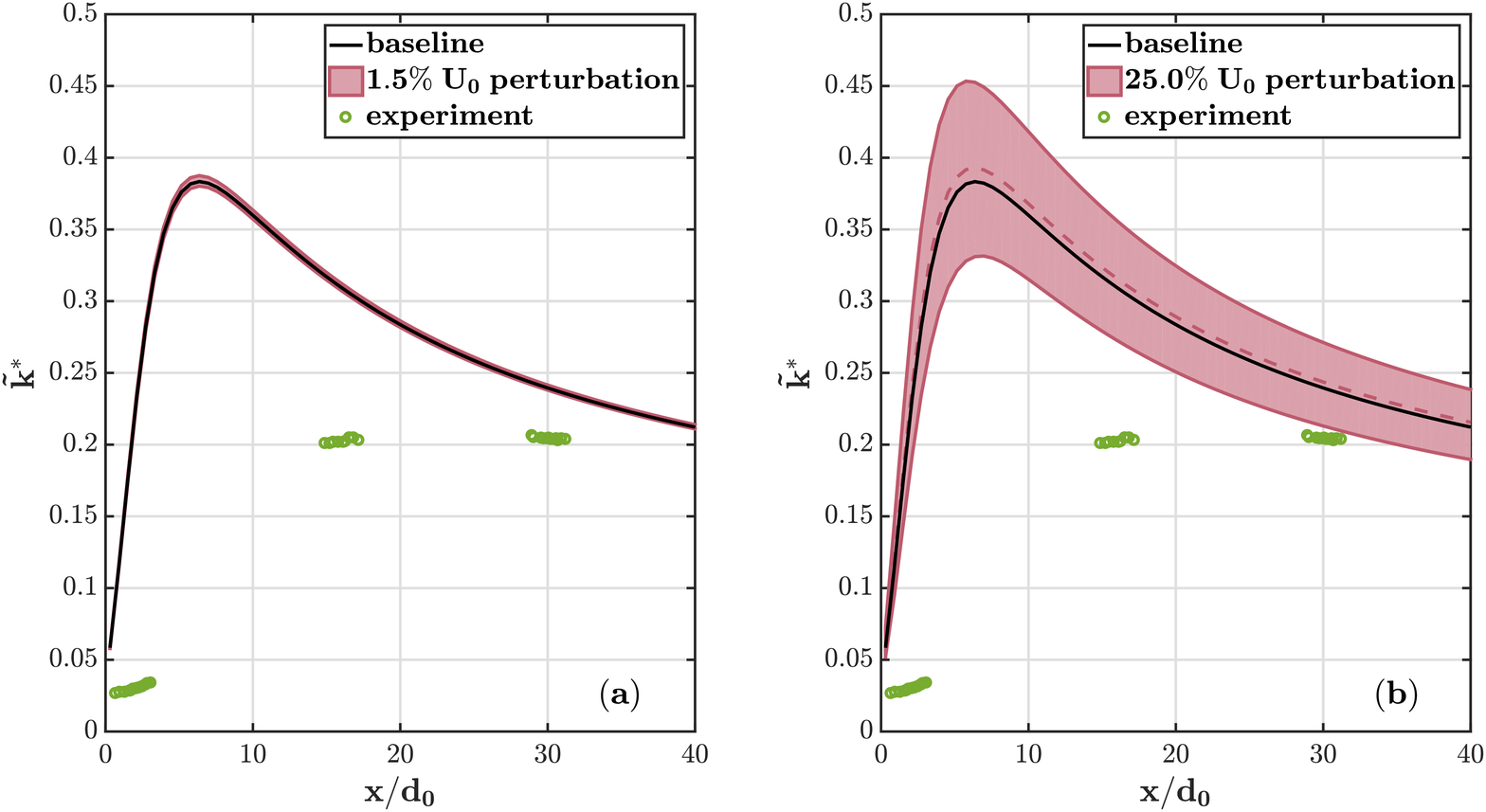}
\caption{Normalized centerline turbulent kinetic energy profile with baseline, estimated uncertainty to $U$ inlet condition, and experimental data. (a) with $1.5\%$ perturbation, (b) with $25.0\%$ perturbation.}
\label{fig:jet-cl-k-U}
\end{center}
\end{figure}

\begin{figure}
\centering
\begin{center}
\includegraphics[width=\textwidth]{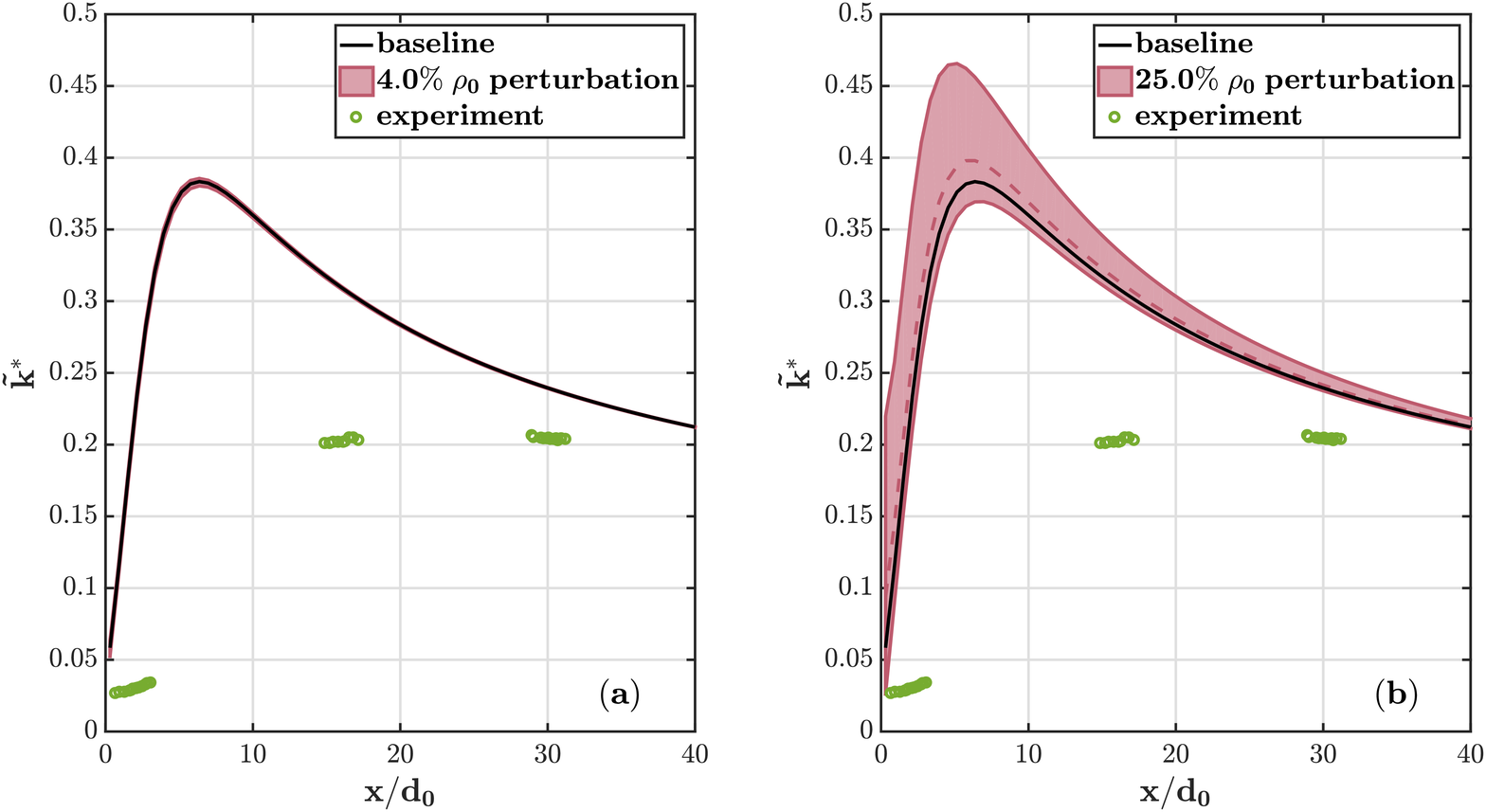}
\caption{Normalized centerline turbulent kinetic energy profile with baseline, estimated uncertainty to $\rho$ inlet condition, and experimental data. (a) with $4.0\%$ perturbation, (b) with $25.0\%$ perturbation.}
\label{fig:jet-cl-k-r}
\end{center}
\end{figure}

Figures \ref{fig:jet-cl-k}-\ref{fig:jet-cl-r} show the normalized centerline velocity profiles with perturbations to the $k$, $U$, and $\rho$ inlet condition, respectively. The centerline coordinate is normalized by the jet nozzle diameter.
The uncertainty bands corresponding to the inlet parameter perturbation cover the estimated $95\%$ interval, and the dashed lines correspond to the expected value. The (a) figures represent the uncertainty reported from the experimental measurements; the (b) figures are based on inflated degrees of inlet uncertainty.
For all three inlet parameters, the sensitivity at experimental conditions is very low. The uncertainty bands barely deviate from the baseline profile and do not match the experimental data, which show higher velocities than the numerical results.
When inflating the variation of the inlet parameters to the $25\%$ level, the estimated $95\%$ intervals become more pronounced, particularly for the $U_{0}$ perturbations. These increased inlet uncertainty bands begin to match the experimental data around $x/d_{0}=30$, while they are still too narrow around $x/d_{0}=16$.

The normalized centerline turbulent kinetic energy can be seen in Figures \ref{fig:jet-cl-k-k}-\ref{fig:jet-cl-k-r}. Again, the sensitivity of the numerical results to perturbations at the $1.5\%$ and $4.0\%$ level is very low. The numerical calculations overpredict the turbulent kinetic energy. Although more pronounced, the estimated uncertainty from the inflated perturbations does not match the experimental results either.

\subsection{Tilted rocket rig case}

\begin{figure}
\centering
\begin{center}
\includegraphics[width=\textwidth]{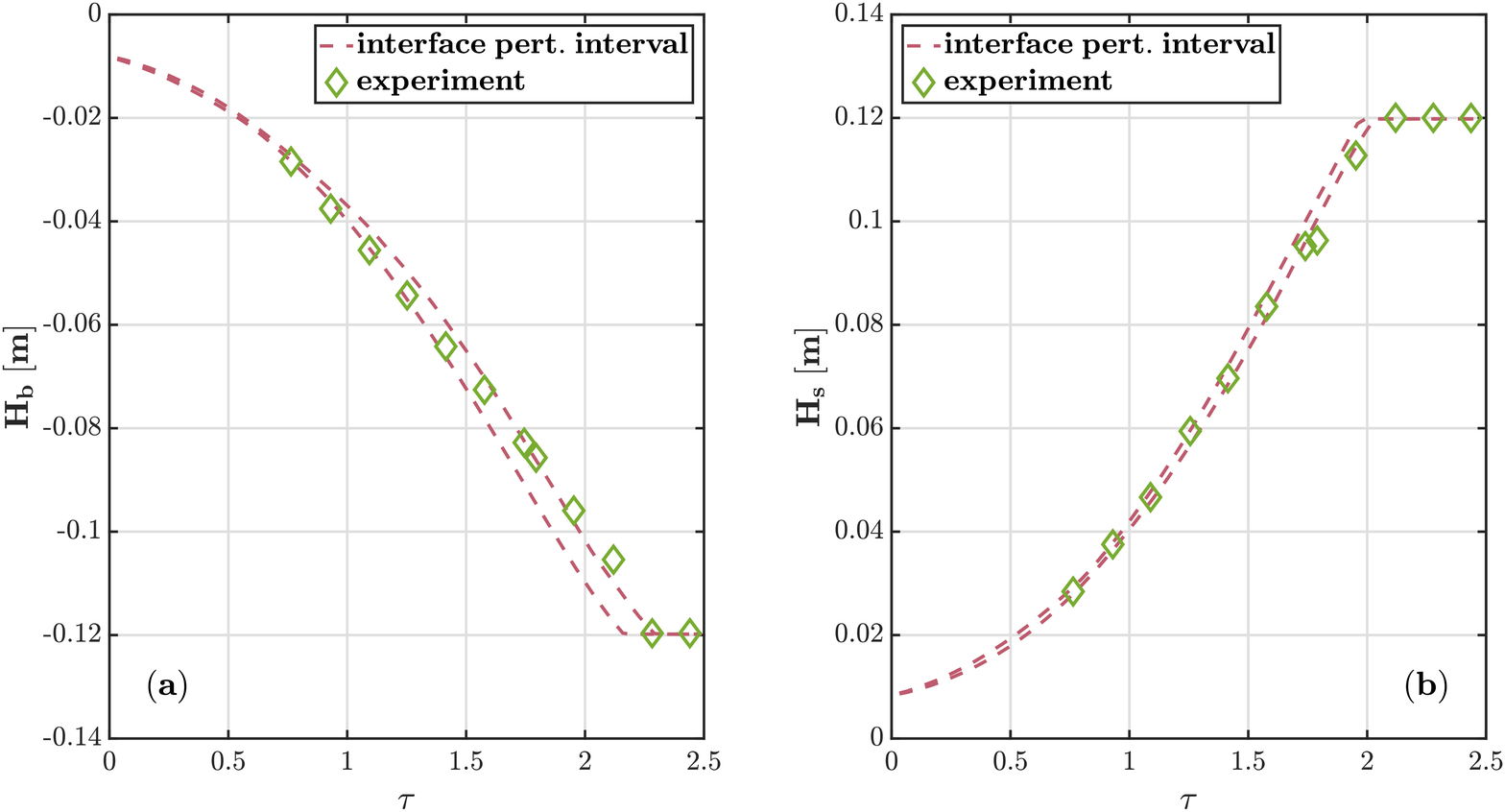}
\caption{Bubble (a) and spike (b) height with $0.4$ mm interface perturbation range, and experimental data.}
\label{fig:tilted-rig-0.4}
\end{center}
\end{figure}

\begin{figure}
\centering
\begin{center}
\includegraphics[width=\textwidth]{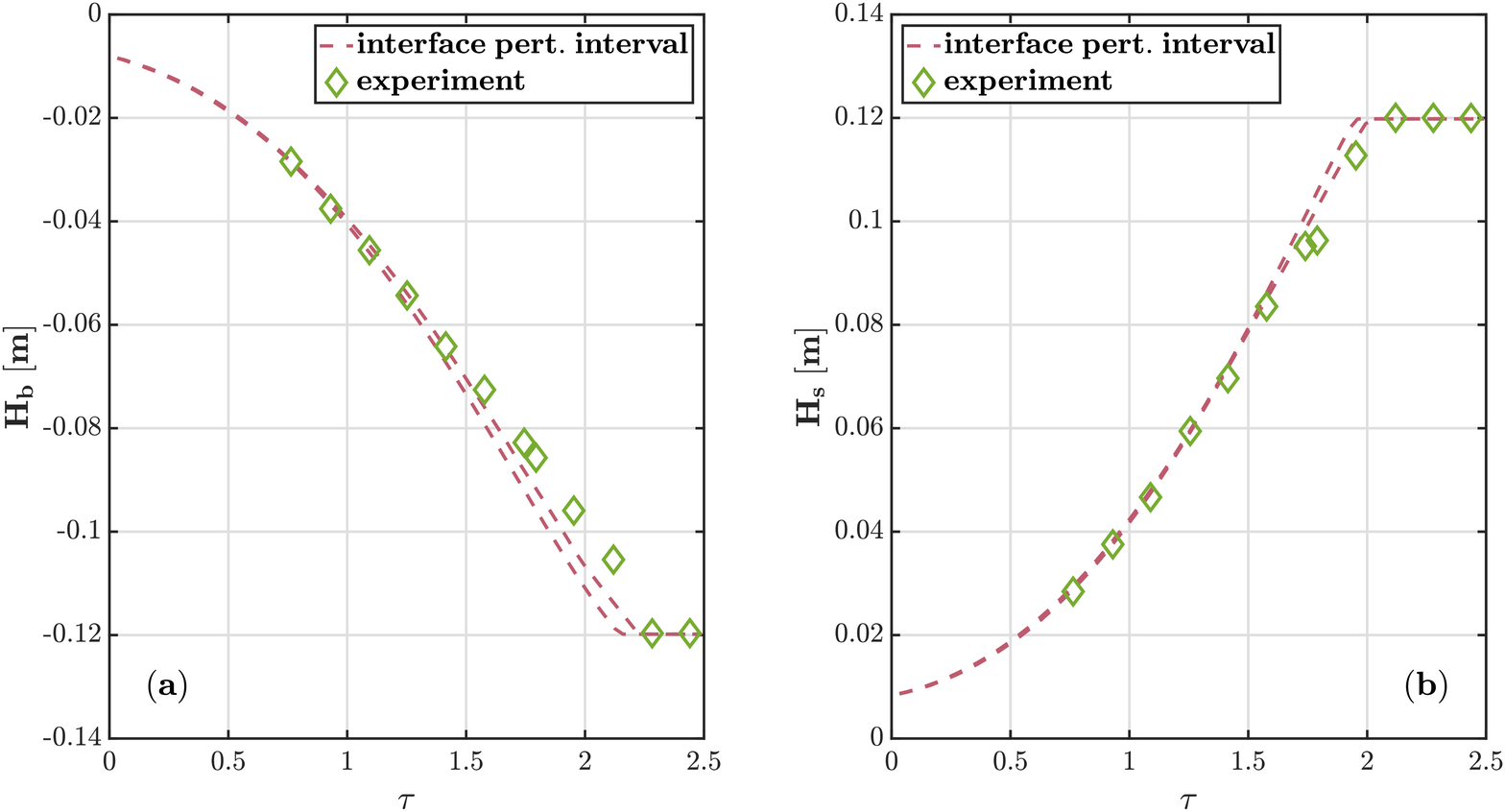}
\caption{Bubble (a) and spike (b) height with $0.2$ mm interface perturbation range, and experimental data.}
\label{fig:tilted-rig-0.2}
\end{center}
\end{figure}

\begin{figure}
\centering
\begin{center}
\includegraphics[width=\textwidth]{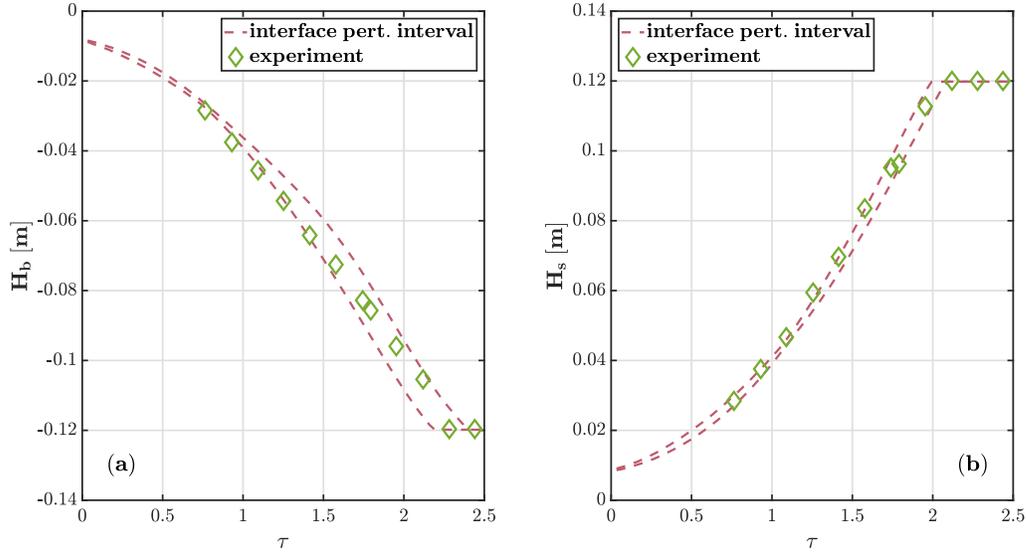}
\caption{Bubble (a) and spike (b) height with $0.6$ mm interface perturbation range, and experimental data.}
\label{fig:tilted-rig-0.6}
\end{center}
\end{figure}

The bubble and spike heights for the three perturbation amplitudes $0.4$ mm, $0.2$ mm, and $0.6$ mm are shown in Figures \ref{fig:tilted-rig-0.4}-\ref{fig:tilted-rig-0.6}. These heights are defined relative to the middle plane at $y=0.12$ m, and therefore they range from $0$ to $-0.12$ m and from $0$ to $0.12$ m, respectively, as can be seen in the figures.

The estimated initial condition uncertainty is significant with respect to the discrepancy between the numerical results and the experimental data. This is a clear difference from the jet in co-flow case.
The uncertainty estimate for the $0.4$ mm case mainly covers the experimental data except for some data points towards the end. With an increased perturbation amplitude of $0.6$ mm, the experimental data is well matched by the wider uncertainty range. Only for the $0.2$ mm case is the uncertainty range too narrow to encompass the experimental data, particularly for the bubble height.

\section{Discussion}

At the jet in co-flow configuration, the results show only little sensitivity to all three inlet parameter perturbations. Even when substantially increasing those perturbations, the estimated uncertainty band does not fully match the experimental results. The numerical centerline velocity predictions are overly diffusive, which is consistent with the overprediction of the turbulent kinetic energy, because an increased $k$ value increases the predicted turbulent mixing.
The model predictions are very robust with respect to the inlet parameter variability, which suggests that estimated accuracy of the experimental measurements is sufficient with respect to the sensitivity of the flow.
There are several possible explanations for the discrepancy between the experimental and the numerical results. One could be an overly simplified setup of the case in the computations. The inlet conditions, for example, are assuming uniform flow for each jet and co-flow inlet. Also, there are confinements in the experimental setup that are not being accounted for in the simulations, such as a bend at the end of the straight section of the turbulent mixing tunnel. One other source of uncertainty is associated with the limitations of RANS simulations and the BHR-1 model.

The tilted rocket rig simulations are significantly more sensitive to the initial conditions than the jet in co-flow simulations are to the inlet conditions. 
The estimated initial condition uncertainty band of the tilted rocket rig matches quite well the experimental data at the nominal perturbation amplitude. 
That perturbation amplitude by itself is a critical parameter, as the results for the $0.2$ mm and $0.6$ mm cases show. For the smaller perturbation, the uncertainty band is much narrower than that for the larger one.
The predictions of this Rayleigh-Taylor instability case hence are less robust to the initial interface perturbations. Therefore, it is important for experiments to characterize the initial state carefully.

For future work, it would be interesting to include estimates of the turbulence model uncertainty. \cite{zhu-2019} used perturbations on the two production terms in the turbulent kinetic energy equation, similar to the eigenspace perturbation framework, to obtain meaningful intervals for variable density flows.
The first two terms on the right-hand side of Eq. (\ref{eq:bhr-transport}) are those production terms. The first one, $a_{j} \frac{\p \bar{p}}{\p x_{j}}$, is related to variable density effects, and it can be perturbed by varying the alignment between the turbulent mass flux velocity and the pressure gradient. The second one, $- R_{ij} \frac{\p \tilde{u}_{i}}{\p x_{j}}$, is related to the shear flow, and the Reynolds stress tensor in it can be perturbed as is done for the eigenspace perturbations.

Furthermore, the investigation of the inlet parameter sensitivity could be extended to the second jet in co-flow setup from \cite{charonko-2017} with the Atwood number $A=0.62$. This jet, unlike the one studied at $A=0.09$, is in the non-Boussinesq regime with a considerable density difference. The more pronounced variable density effects could affect the inlet parameter sensitivity.

\section*{Acknowledgments} 

This work is funded by a grant from Los Alamos National Laboratory.
The authors acknowledge use of computational resources from the Certainty cluster awarded by the National Science Foundation to CTR. 

\bibliographystyle{ctr}


\begin{thebibliography}{} 




\bibitem[Andrews(2013)]{andrews-2013}
\textsc{Andrews, M.~J.} 2013
The use of dual-number-automatic-differentiation with sensitivity analysis to investigate physical models.
\emph{J. Fluids Eng.} \textbf{135}, 061206.

\bibitem[Andrews {\em et al.}(2014)]{andrews-2014}
\textsc{Andrews, M.~J., Youngs, D.~L., Livescu, D. \& Wei, T.} 2014
Computational studies of two-dimensional Rayleigh-Taylor driven mixing for a tilted-rig.
\emph{J. Fluids Eng.} \textbf{136}, 091212.

\bibitem[Besnard {\em et al.}(1992)]{besnard-1992}
\textsc{Besnard, D., Harlow, F.~H., Rauenzahn, R.~M. \& Zemach, C.} 1992
Turbulence transport equations for variable-density turbulence and their relationship to two-field models.
\emph{LANL Report LA-12303-MS}.

\bibitem[Boersma(1998)]{boersma-1998}
\textsc{Boersma, B.~J., Brethouwer, G. \& Nieuwstadt, F.~T.~M.} 1998
A numerical investigation on the effect of the inflow conditions on the self-similar region of a round jet
\emph{Phys. Fluids} \textbf{10}, 899-909.

\bibitem[Brown {\em et al.}(2015)]{brown-2015}
\textsc{Brown, J., Sarno-Smith, L., Israel, D. \& Denissen, N.} 2015
Turbulence Modeling with BHR in OpenFOAM.
\emph{2015 Final Reports from the LANL Computational Physics Student Summer Workshop}, 186-204.

\bibitem[Charonko \& Prestridge(2017)]{charonko-2017}
\textsc{Charonko, J.~J. \& Prestridge, K.} 2017
Variable-density mixing in turbulent jets with coflow
\emph{J. Fluid Mech.} \textbf{825}, 887-921.

\bibitem[Charonko \& Vlachos(2013)]{charonko-2013}
\textsc{Charonko, J.~J. \& Vlachos, P.~P.} 2013
Estimation of uncertainty bounds for individual particle image velocimetry measurements from cross-correlation peak ratio.
\emph{Measurement Science and Technology} \textbf{24}, 065301.

\bibitem[Denissen {\em et al.}(2014)]{denissen-2014}
\textsc{Denissen, N.~A., Rollin, B., Reisner, J.~M. \& Andrews, M.~J} 2014
The tilted rocket rig: a Rayleigh-Taylor test case for RANS models.
\emph{J. Fluids Eng.} \textbf{136}, 091301.

\bibitem[Dimonte \& Schneider(2000)]{dimonte-2000}
\textsc{Dimonte, G. \& Schneider, M.} 2000
Density ratio dependence of Rayleigh-Taylor mixing for sustained and impulsive acceleration histories.
\emph{Phys. Fluids} \textbf{12}, 304-321.

\bibitem[Emory {\em et al.}(2013)]{emory-2013}
\textsc{Emory, M., Larsson, J. \& Iaccarino, G.} 2013
Modeling of structural uncertainties in Reynolds-averaged Navier-Stokes closures.
\emph{Phys. Fluids} \textbf{25}, 110822.

\bibitem[Granados-Ortiz {\em et al.}(2018)]{granado-ortiz-2018}
\textsc{Granados-Ortiz, F.-J., Arroyo, C.~P., Puigt, G., Lai, C.-H. \& Airiau, C.} 2018
On the influence of uncertainty in computational simulations of high speed jet flow from an aircraft exhaust.
\emph{Computers \& Fluids} \textbf{138}, 139-158.

\bibitem[G{\"u}rkan {\em et al.}(2011)]{gurkan-2011}
\textsc{G{\"u}rkan, S., Gernaey, K.~V., Neumann, M.~B., van Loosdrecht, M.~C.~M. \& Gujer, W.} 2011
Global sensitivity analysis in wastewater treatment plant model applications: Prioritizing sources of uncertainty
\emph{Water Research} \textbf{45}, 639-651.

\bibitem[Iaccarino {\em et al.}(2017)]{iaccarino-2017}
\textsc{Iaccarino, G., Mishra, A.~A. \& Ghili, S.} 2017
Eigenspace perturbations for uncertainty estimation of single point turbulence closures.
\emph{Phys. Rev. Fluids} \textbf{2}, 024605.

\bibitem[Kucherenko {\em et al.}(1996)]{kucherenko-1996}
\textsc{Kucherenko, Y.~A., Balabin, S.~I., Cherret, R. \& Haas, J.~F.} 1996
Experimental Investigation into inertial properties of Rayleigh-Taylor turbulence.
\emph{Laser and Particle Beams} \textbf{15}, 25-31.

\bibitem[Smeeton \& Youngs(1987)]{smeeton-1987}
\textsc{Smeeton, V.~S. \& Youngs, D.~L.} 1987
Experimental investigation of turbulent mixing by Rayleigh-Taylor instability, III.
\emph{AWE Report No. O-35/87}.

\bibitem[Youngs(1989)]{youngs-1989}
\textsc{Youngs, D.~L.} 1989
Modelling turbulent mixing by Rayleigh-Taylor instability.
\emph{Physica D: Nonlinear Phenomena} \textbf{37}, 270-287.

\bibitem[Youngs(2009)]{youngs-2009}
\textsc{Youngs, D.~L.} 2009
Application of monotone integrated large eddy simulation to Rayleigh–Taylor mixing.
\emph{Phil. Trans. R. Soc. A} \textbf{367}, 2971-2983.

\bibitem[Zhu {\em et al.}(2019)]{zhu-2019}
\textsc{Zhu, H., Mishra, A.~A., Hayes, J. \& Iaccarino, G.} 2019
Quantifying the structural uncertainty in the BHR model for variable density flows.
\emph{Annual Research Briefs}, Center for Turbulence Research, Stanford University.













\end{thebibliography}

\end{document}